# Refinement of genetic variants needs attention

Omar Abdelwahab[1,2,3,4], Davoud Torkamaneh[1,2,3,4]*

[1]*Département de Phytologie, Université Laval, Québec, Canada*
[2]*Institut de Biologie Intégrative et des Systèmes (IBIS), Université Laval, Québec, Canada*
[3]*Centre de recherche et d'innovation sur les végétaux (CRIV), Université Laval, Québec, Canada*
[4]*Institut intelligence et données (IID), Université Laval, Québec, Canada*

* Corresponding author: davoud.torkamaneh.1@ulaval.ca

## Abstract

Variant calling refinement is crucial for distinguishing true genetic variants from technical artifacts in high-throughput sequencing data. Manual review is time-consuming while heuristic filtering often lacks optimal solutions. Traditional variant calling methods often struggle with accuracy, especially in regions of low read coverage, leading to false-positive or false-negative calls. Here, we introduce VariantTransformer, a Transformer-based deep learning model, designed to automate variant calling refinement directly from VCF files in low-coverage data (10-15X). VariantTransformer, trained on two million variants, including SNPs and short InDels, from low-coverage sequencing data, achieved an accuracy of 89.26% and a ROC AUC of 0.88. When integrated into conventional variant calling pipelines, VariantTransformer outperformed traditional heuristic filters and approached the performance of state-of-the-art AI-based variant callers like DeepVariant. Comparative analysis demonstrated VariantTransformer's superiority in functionality, variant type coverage, training size, and input data type. VariantTransformer represents a significant advancement in variant calling refinement for low-coverage genomic studies.

# Introduction

Genetic variants are considered the backbone for identifying mutations within populations and are fundamental to genetic screening tools [1]. Variant calling is the process of identifying differences between an individual's genome and a reference genome, encompassing single nucleotide polymorphisms (SNPs), small insertions and deletions (InDels), and structural variations [2]. However, the raw output from high-throughput sequencing data often requires refinement to distinguish true genetic variants from technical artifacts. Variant calling refinement, a critical post-processing stage following variant calling, is aimed at filtering out erroneous calls and enhancing the accuracy of detected variants. While variant calling provides an initial catalog of genomic variations, refinement is essential to minimize false positives and false negatives, ensuring that downstream analyses are built upon a foundation of reliable genetic data [3,4].

Typically, variant refinement involves heuristic filtering and/or manual inspection. Heuristic filtering entails establishing project-specific thresholds for key metrics such as read depth, variant allele fraction (VAF), base quality, read quality, and mapping quality scores [5]. Manual review, while time-intensive, enhances confidence in specific variants by uncovering patterns typically overlooked by conventional variant callers through direct visual inspections of the variants using genomic viewers like Integrative Genomic Viewer (IGV) [6,7]. Although heuristic filtering offers speed, it often fails to provide the optimal solution if it does not align with user-defined thresholds [8–11]. Strict heuristic filtering leads to more false negatives, while relaxed filtering results in a higher number of false positives. Despite its importance, the refinement process in variant calling is often underdeveloped and lacks comprehensive representation in genomic workflows.

With the emergence of artificial intelligence (AI), new models have been introduced to enhance the process of variant calling refinement [12–14]. These early efforts have shown potential in utilizing machine and deep learning (ML and DL) techniques to improve the precision of variant calling. However, they often focused on limited datasets predominantly associated with cancer-related diseases and are derived from high-coverage sequencing data, restricting their utility across a broader spectrum of genomic studies[15].

Transformers, a revolutionary class of deep learning models originally developed for natural language processing (NLP), have demonstrated exceptional capabilities in identifying complex patterns and dependencies in sequential data [16]. Their unique architecture, which includes "self-attention" and "feed-forward neural network" layers allows for dynamic learning of correlations among features, thereby enhancing classification tasks [16]. Given their adaptability, Transformers are ideally suited to address the challenges of variant calling refinement, enabling the thorough analysis of extensive genomic datasets and facilitating the extraction of high-quality genomic variations.

In this study, we introduce VariantTransformer, a model that automates the variant calling refinement process. This model is capable of directly processing variant calling format files (VCFs) and can be seamlessly integrated into standard variant calling pipelines such as BCFTools [17] and GATK [18]. Specifically designed for low-coverage sequencing data, VariantTransformer uses a dataset of 2 million variants, the largest utilized for such a purpose to date. We treat each variant in the VCF as a distinct "sentence", thus conceptualizing the refinement process as a "sentence classification problem" where the goal is to categorize each variant as "Pass" or "Fail". This classification is then used to update the "Filter" column in the VCF, thereby enhancing the model's applicability and utility in genomic studies.

## Methods

### Sequencing data

The FASTQ files containing sequencing data, generated on an Illumina HiSeq2500, for three samples (HG003, HG006, and HG007) with sequence coverages of 10.5X, 13.6X, and 12.6X respectively, were procured from the Genome in a Bottle (GIAB) Consortium [19], accessed via the NIST GIAB FTP site (https://ftp-trace.ncbi.nlm.nih.gov/giab/ftp/data/). We used SAMtools [20] to determine the coverage of samples, utilizing the '-a' option, to consider all positions within the reads. The alignment of the raw FASTQ files to the human reference genome GRCh38 (GCA_000001405.15_GRCh38_no_alt_analysis_set.fna) was performed using Sentieon BWA-MEM [21].

### Training and testing Data

Variant calling on the aligned BAM files was performed with GATK4 HaplotypeCaller [18] and BCFTools [17]. To ensure accurate variant classification, we used the latest GIAB truth sets v4.2.1 [22] to update the "FILTER" column in the VCFs; variants matching the truth sets were labeled "PASS" while all others were labeled as "FAIL".

During preprocessing, we simplified the dataset by removing non-essential columns ("Chrom," "POS," "REF," and "ALT") and consolidating the remaining data into a single column. This restructuring facilitated the transformation of the dataset into a sentence classification format, where the merged column represented the 'sentences' and the "FILTER" column the target labels.

For training purposes, we merged the updated VCFs (generated from GATK4 and BCFTools) from the sample HG003. To reduce computational cost in the training process, we randomly selected 2 million variants from the merged VCF for initial model training and validation. Table 1 demonstrates the number of variants generated at each step. The selected dataset was then split into 60% for training and 40% for validation, with the remaining variants of the HG003 and the other VCFs from the other samples being used for further testing.

**Table 1**. Breakdown of "PASS" and "FAIL" variants from sample HG003, processed with GATK4 and BCFTools.

| HG003 sample | PASS | FAIL | Total |
| --- | --- | --- | --- |
| BCFTools | 3,714,910 | 708,823 | 4,423,733 |
| GATK4 | 3,668,708 | 964,363 | 4,633,071 |
| Total_merged_file | 7,383,618 | 1,673,186 | 9,056,804 |
| Training and validation set | 1,630,866 | 369,134 | 2,000,000 |

In the testing process, we used two other samples (HG006 and HG007) (Table 2). We tested in batches of 10,000 to generate probabilities for further model performance analyses.

**Table 2**. Variant counts obtained from samples HG006 and HG007 via GATK4 and BCFTools, detailing "PASS" and "FAIL" variants.

| Sample | Coverage (X) | Variant Caller | PASS | FAIL | Total |
| --- | --- | --- | --- | --- | --- |
| HG006 | 13.6 | BCFTools | 3,694,662 | 826,173 | 4,520,835 |
| | | GATK4 | 3,666,347 | 1,066,280 | 4,732,627 |
| HG007 | 12.6 | BCFTools | 3,685,056 | 847,135 | 4,532,191 |
| | | GATK4 | 3,650,142 | 1,042,385 | 4,692,527 |

## Model development and analysis

A DL model was developed based on the Transformers architecture [16] to automate the variant calling refinement process. We used the "BertForSequenceClassification" model from Hugging Face [23] while tuning some of the parameters in the configuration and tokenization steps to achieve better performance (Supplementary Table S1). This selection was made due to the model's aptitude for handling sequence-based data with complex dependencies. Parameters were finely tuned, including the addition of variant calling specific vocabulary to the BertTokenizer and adjustments to several hyperparameters (Supplementary Table S3) to optimize the model's predictive accuracy and computational efficiency. For the rest of the parameters, we used the default values mentioned in the Hugging Face documentation [23]. The model trained over 21 epochs with a batch size of 1,300 using the AdamW optimizer [24], focusing on balancing performance and resource utilization.

## Evaluation Metrics

To thoroughly evaluate model performance, we employed several accuracy metrics including the AUC (Area Under The Curve) ROC (Receiver Operating Characteristics) curve [25], Matthews correlation coefficient (MCC) [26–29], accuracy, precision, recall, and F1 score [30]. For further model evaluation, we reported MCC, also known as the phi coefficient, where a coefficient of +1 indicates an ideal prediction, 0 signifies an average random prediction, and -1 denotes a reverse prediction [27,29]. All accuracy metrics were generated using scikit-learn [31]. Moreover, we compared the model performance to default filtering parameters of conventional variant callers (BCFTools and GATK4), AI-based variant caller (DeepVariant [32]), and AI-based tool for refinement of somatic variant calling (DeepSVR [14]) considering both PASS and FAIL variants. For BCFTools and GATK4, we integrated VariantTransformer into each pipeline and compared the performance of the model against default filters. For GATK4, the default filters were QD < 2.0, FS > 60.0, MQ < 40.0, SOR > 4.0, MQRankSum < -12.5, and ReadPosRankSum < -8.0. For BCFTools, we applied only the QUAL>=20 filter. As for DeepVariant, we compared the performance of each model-integrated conventional variant caller against DeepVariant. DeepSVR was compared against VariantTransformer in terms of data preparation, multiple accuracy metrics, computational complexity, and user experience. All plots were generated using the MatPlotlib library [33] or ggplot2 [34].

## Data availability

The complete dataset utilized in this study is accessible via the NIST GIAB FTP site: https://ftp-trace.ncbi.nlm.nih.gov/giab/ftp/data/. All the resources related to the analyses, including preprocessing code, training and testing code, and the trained model can be found in the GitHub repository: https://github.com/Omar-Abd-Elwahab/VCF_filter. This repository also offers guidance on using the model for filtering VCFs, adapting it to other sequencing technologies, or integrating it into different variant calling pipelines.

## Results

### Model development

VariantTransformer was developed using a dataset of 2 million variants, including both NPs and InDels, sourced from the GIAB sample HG003. The variants were called using GATK4 and BCFTools. Of the two million variants, 1,630,866 matched the latest GIAB truth sets and were classified as "PASS", while the remaining 369,134 were classified as "FAIL". To avoid overfitting, the data was randomly split into 60% for training and 40% for validation. The model achieved an accuracy of 89.26% and ROC AUC score of 0.88, demonstrating robust performance (Figure 1a).

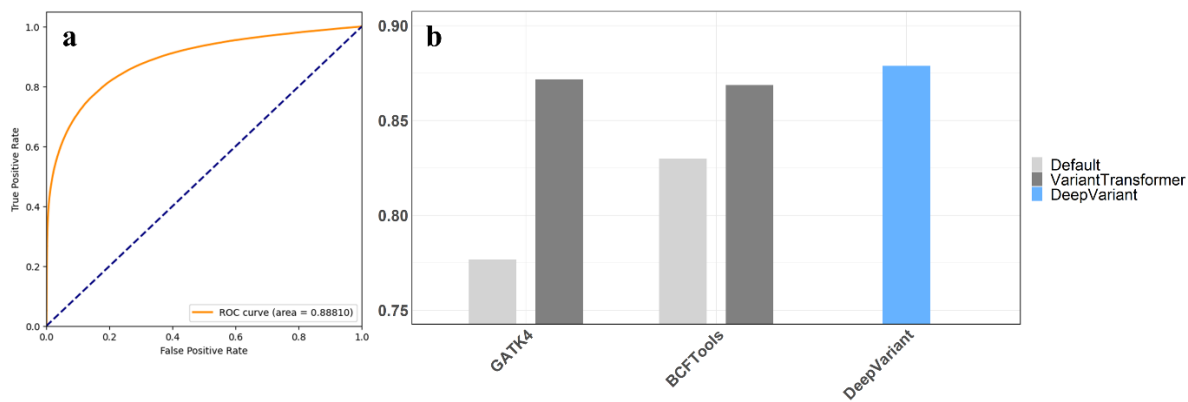

Figure1. a) ROC AUC curve for the trained model. b) VariantTransformer's average performance in refining variants called from the GIAB samples HG006 and HG007, benchmarked against default filtering and DeepVariant.

## Model testing and evaluation

To test VariantTransformer, we integrated it into two conventional variant calling pipelines (BCFTools and GATK4). We processed the two GIAB samples HG006 and HG007 through these pipelines and subsequently applied VariantTransformer to refine the variant calls. The model significantly outperformed the default threshold-based filtering, achieving an overall accuracy of 87%, compared to 78% for GATK4 and 83% for BCFTools, thus approaching the performance of the AI-based variant caller, DeepVariant, which has an accuracy of 88% (Figure 1b).

During the testing phase, batches of 10,000 variants were processed to generate performance metrics such as MCC scores, ROC AUC scores, accuracy, precision, recall, and F1score. The aggregated results, outlined in Table 4, provide a comprehensive evaluation of VariantTransformer's efficacy when integrated into the aforementioned variant calling pipelines.

Table 4. Performance of VariantTransformer across different metrics in the variant calling pipelines for samples HG006 and HG007.

| Sample | Coverage (X) | Variant Caller | Accuracy* | Refined Accuracy** | Precision | Recall | F1score | ROC AUC | MCC |
|---|---|---|---|---|---|---|---|---|---|
| HG006 | 13.6 | BCFTools | 83.31% | **87.03%** | 0.85957 | 0.87027 | 0.85845 | 0.86413 | 0.51194 |
|  |  | GATK4 | 77.50% | **87.38%** | 0.87004 | 0.87375 | 0.86385 | 0.88876 | 0.60805 |
| HG007 | 12.6 | BCFTools | 82.63% | **86.68%** | 0.85635 | 0.86678 | 0.85197 | 0.83063 | 0.50117 |
|  |  | GATK4 | 77.82% | **86.94%** | 0.86646 | 0.86944 | 0.85692 | 0.85605 | 0.58638 |

\* Default filtering

\*\* After VariantTransformer refinement

## Comparative analysis: assessing performance against existing pipelines and models

In contrast to most performance analysis studies that use the hap.py tool [35] and focus solely on variants passing the filters, our study, comprehensively assessed all variants, including those identified as "FAIL". This comprehensive approach considers all four parameters: true positives (TP), false positives (FP), true negatives (TN), and false negatives (FN).

The performance of VariantTransformer was compared with BCFTools and GATK4 pipelines, and with DeepVariant, focusing on two GIAB samples (HG006 and HG007) with coverage of 13.6X and 12.6X (Table 5). For sample HG006 using BCFTools, from 4,520,835 variants, VariantTransformer identified 3,993,987 as "PASS" with an accuracy of 87.03%. When using GATK4 on the same sample, the model called 4,027,338 as "PASS" with an accuracy of 87.38%. For sample HG007 using BCFTools, VariantTransformer identified 4,044,040 as "PASS" with an accuracy of 86.68%. When using GATK4 on HG007, the model called 4,060,472 as "PASS" with an accuracy of 86.94%. Comparatively, DeepVariant demonstrated slightly higher overall accuracy, with 87.95% for HG006 and 87.78% for HG007. However, the number of FP was notably higher for DeepVariant compared to VariantTransformer, highlighting our model's balance between accuracy and precision.

Table 5. Detailed comparison of VariantTransformer's performance when integrated with GATK4 and BCFTools against DeepVariant

| Sample | Coverage (X) | Variant Caller | Input* | PASS** | TP | FP | FN | TN | Accuracy |
|---|---|---|---|---|---|---|---|---|---|
| HG006 | 13.6 | BCFTools | 4,520,835 | 3,993,987 | 3,551,084 | 442,903 | 143,576 | 383,270 | **87.03%** |
| | | GATK4 | 4,732,627 | 4,027,338 | 3,548,098 | 479,240 | 118,249 | 587,039 | **87.38%** |
| | | DeepVariant | 6,643,643 | 4,213,736 | 3,593,719 | 620,018 | 180,668 | 2,249,241 | **87.95%** |
| HG007 | 12.6 | BCFTools | 4,532,191 | 4,044,040 | 3,562,660 | 481,380 | 122,394 | 365,755 | **86.68%** |
| | | GATK4 | 4,692,527 | 4,060,472 | 3,548,979 | 511,493 | 101,163 | 530,891 | **86.94%** |
| | | DeepVariant | 6,716,326 | 4,247,514 | 3,597,227 | 650,288 | 170,251 | 2,298,563 | **87.78%** |

\* Total called variants
\*\* Identified as PASS by VariantTransformer in case of BCFTools and GATK4, or Identified as PASS by DeepVariant filter.

## Discussion

The development and validation of the Transformer-based VariantTransformer model, for the refinement of variant calling in low-coverage genomic data, represent a significant advance in leveraging AI technologies in genomics. This model demonstrates an impressive variant refinement accuracy of 89.26%, which not only outperforms conventional refinement methods but also aligns closely with contemporary AI-based tools. The higher prediction performance demonstrates that this automated strategy has the potential to substitute the conventional threshold-based filtering methods, particularly in contexts involving low-coverage genomic data.

VariantTransformer's effectiveness in processing —a technique borrowed from NLP [16] — highlights its ability to handle the complex patterns inherent in genomic sequence data. By interpreting these patterns, the model distinguishes true genetic variants from technical artifacts [36] with high efficiency, achieving ROC AUC scores that affirm its capacity to differentiate variant classes across all thresholds, thus enhancing its utility in varied analytical scenarios.

The model's success can be attributed to several key factors. First, the use of an extensive dataset comprising 2 million variants, including both SNPs and short InDels, provided a robust foundation for both training and validation. Second, modifications to the default BERT model parameters, specifically, reductions in hidden size and the number of attention heads, have tailored the model to handle the specific complexity of genomic data while optimizing computational efficiency [16]. Finally, the MCC values highlight the model's balanced accuracy, considering both positive and negative classes, which is essential for applications in genomic studies where both sensitivity and specificity are critical [28]. The ROC AUC scores further affirm the model's exceptional capability to distinguish between the variant classes across all thresholds, emphasizing its effectiveness in various scenarios [37].

When comparing VariantTransformer to DeepVariant, it is noteworthy that despite DeepVariant's slightly higher overall accuracy in some cases (87.95% for HG006 and 87.78% for HG007 with a difference of less than 1% when compared to VariantTransformer), it also presented a significantly higher rate of FPs for both samples. For sample HG006, DeepVariant had approximately 40% more FPs than BCFTools and approximately 29.38% more FPs than GATK4. For sample HG007, DeepVariant had approximately 35.07% more FPs than BCFTools and approximately 27.11% more FPs than GATK4. This comparison underscores VariantTransformer's efficiency in minimizing incorrect variant identifications—a crucial advantage in genomic analytics.

Furthermore, the comparison with DeepSVR [14], a recently developed deep learning model to automate variant calling refinement, reveals significant distinctions[14]. DeepSVR classified SNPs into "Pass", "Fail", or "Ambiguous" with Fliess' Kappa statistics [38] indicating fair agreement [14]. DeepSVR is designed and trained with a small training set (41,000 variants) specifically for deep coverage (300-1000X) cancer-type samples, processes only SNPs, and requires extensive preprocessing of data [14]. In contrast, VariantTransformer supports both SNPs and InDels and does not require data preprocessing, allowing it to operate directly with VCF files from standard variant calling pipelines like BCFTools and GATK4 (a detailed comparison of DeepSVR and VariantTransformer is presented in Supplementary Table S2). This adaptability makes VariantTransformer a more versatile tool suitable for a broader range of genomic studies.

However, there are limitations to our approach. The training dataset, derived mainly from well-characterized genomic regions provided by the GIAB consortium, might not fully represent the diversity of genomic scenarios encountered in wider research and clinical contexts. Moreover, the model configuration is primarily optimized for data generated from Illumina platforms. Future studies could expand the model's training scope to include more diverse and challenging genomic landscapes, potentially enhancing its applicability and accuracy in a wider array of genomic settings.

# Conclusion

VariantTransformer represents a significant advancement in the field of genomic studies, particularly in the refinement variant calling. Achieving an accuracy of 89.26%, this Transformer-based model surpasses traditional methods and rivals existing AI-based tools in performance. By conceptualizing variant calling as a text classification problem, VariantTransformer efficiently differentiates between true variants and technical artifacts showcasing the robust application of NLP techniques in genomics. The demonstrated capabilities of VariantTransformer suggest its potential to substitute conventional threshold-based filtering of low-coverage genomic data, offering a more accurate and efficient alternative. While the results are promising, the scope for further enhancements remains vast. Overall, this study highlights the feasibility of using Transformer models within the realm of genomics and underscores the transformative potential of AI technologies in enhancing the accuracy and efficiency of variant calling, particularly in low-coverage genomic studies, thus paving the way for significant advancements in the field.

## Acknowledgments

The authors wish to thank Génome Québec, Genome Canada, the government of Canada, the Ministère de l'Économie et de l'Innovation du Québec, the Canadian Field Crop Research Alliance, Semences Prograin Inc., Sollio Agriculture, Grain Farmers of Ontario, Barley Council of Canada, and Université Laval. The authors wish also to thank GIAB for providing the reference variants of the samples.

## Funding

This work was funded by Genome Canada [#6548] under Genomic Applications Partnership Program (GAPP).

## Competing Interests

The authors declare that the research was conducted in the absence of any commercial or financial relationships that could be construed as a potential conflict of interest.

## Authors' Contributions

Conceptualization, O.A. and D.T.; data curation, model development and formal analysis, O.A.; resources, D.T.; writing, review and editing, O.A., and D.T.; supervision, D.T.; project administration, D.T.; funding acquisition, D.T. All authors have read and agreed to the published version of the manuscript.

**Supplementary Table S1**. Training parameters employed during model development and final parameters used in the fine-tuned model. Default refers to pre-defined model parameters from Hugging Face documentation, while Model represents tuned parameters.

| Parameters | Default | VariantTransformer |
|---|---|---|
| **Vocab_size** | 30,522 | 30,567 |
| **Never_split** | False | VCF unique words |
| **Hidden_size** | 768 | 256 |
| **num_hidden_layers** | 12 | 8 |
| **num_attention_heads** | 12 | 8 |
| **intermediate_size** | 3072 | 512 |
| **max_position_embeddings** | 512 | 256 |
| **classifier_dropout** | Optional (zero) | 0.1 |

**Supplementary Table S2.** Comparison between DeepSVR and VariantTransformer models.

| Model | DeepSVR | VariantTransformer |
|---|---|---|
| General functionality | Filters SNPs from specific cancer types samples | Filters SNPs and InDels from any raw VCF file with no preprocessing |
| Variant type | SNPs only | SNPs and InDels |
| Training size | 41,000 variants | 2,000,000 variants |
| Model's spectrum | Limited cancer types | Any sample |

| Training data preparation | Manual review of a small number of variants from cancer cases | Benchmark GIAB truth sets |
| Input data type | Requires preprocessing to a specific format with +50 features | No preprocessing: Any VCF file produced from a conventional variant calling pipeline (i.e. BCFTools and GATK4) |
| Testing data depth of coverage | Deep coverage of 312X up to 1,000X for some cases | Low coverage of 10X to 13.6X |